\documentclass{optica-article}

\journal{opticajournal} 

\articletype{Research Article}

\usepackage{lineno}
\usepackage{makecell}

\begin{document}

\title{Semiconductor-on-diamond cavities for spin optomechanics}

\author{Xinyuan Ma\authormark{1}, Prasoon K. Shandilya\authormark{1} and Paul E. Barclay\authormark{1,*}}

\address{\authormark{1}Department of Physics and Astronomy and Institute for Quantum Science and Technology, University of Calgary, 2500 University Drive NW, Calgary, AB T2T 4E4, Canada\\
}
\email{\authormark{*}pbarclay@ucalgary.ca} 


\begin{abstract*}
Optomechanical cavities are powerful tools for classical and quantum information processing that can be realized using nanophotonic structures that co-localize optical and mechanical resonances. Typically, phononic localization requires suspended devices that forbid vertical leakage of mechanical energy. Achieving this in some promising quantum photonic materials such as diamond requires non-standard nanofabrication techniques, while hindering integration with other components and exacerbating heating related challenges. As an alternative, we have developed a semiconductor-on-diamond platform that co-localizes phononic and photonic modes without requiring undercutting. We have designed an optomechanical crystal cavity that combines high optomechanical coupling with low dissipation, and we show that this platform will enable optomechanical coupling to spin qubits in the diamond substrate. These properties demonstrate the promise of this platform for realizing quantum information processing devices based on spin, phonon, and photon interactions.
\end{abstract*}

\section{Introduction}

Interfaces between light and other quantum systems are central to applications in quantum networking \cite{Kimble2008}, sensing \cite{degen2017quantum, Li2021}, and computing \cite{Ladd2010}. Optomechanical devices can play an essential role in realizing quantum interfaces by harnessing the sensitivity of mechanical resonators to a wide range of physical forces and fields \cite{Safavi2019}. Optomechanical cavities, which confine light near or within a mechanical resonator, allow the resonator's dynamics to be sensitively monitored \cite{Aspelmeyer2014} and the properties of systems interacting with it to be probed. These devices also allow light to control the state of mechanical resonators and other systems interacting with them. For example, coherent coupling between photons and phonons has enabled demonstrations of transducers between light and superconducting qubits and microwaves \cite{Bochmann2013, Mirhosseini2020, Jiang2020, Delaney2022} and electron spin qubits \cite{Shandilya2021}. 

Key to operation of cavity optomechanical devices is the strength of the coupling between their optical and mechanical resonances relative to their photonic and phononic dissipation rates \cite{Safavi2019}. These properties can be optimized through device engineering, as demonstrated in optomechanical crystal cavities \cite{Eichenfield2009c} and whispering gallery mode resonators \cite{Schliesser2010}. Optomechanical coupling in these devices is enhanced by localizing optical and mechanical fields within overlapping volumes. In these and most other chip-based cavity optomechanical devices,  mechanical and optical dissipation is reduced by suspending the device, preventing photons and phonons from leaking into the substrate supporting the device. However, suspended devices have several disadvantages. Poor thermalization of suspended membranes limits their performance in quantum optomechanical applications \cite{Meenehan2015}, and suspended structures are challenging to integrate with other photonic components. In addition, fabrication of suspended structures is difficult in material systems that are not readily available in thin film form, such as diamond and other crystals.

Realising non-suspended cavity optomechanical devices from diamond is of interest because of their potential as a platform for connecting electron spin qubits associated with defects in the diamond carbon lattice to light and other quantum systems \cite{Lee2017,Shandilya2021,Wang2020}. Diamond electron spin qubits are one of the most advanced solid state systems for realizing quantum memories and quantum networks \cite{atature2018material,shandilya2022diamond}. Although recent nanofabrication advances have led to demonstrations of suspended diamond structures \cite{Khanaliloo2015,Burek2016}, these devices suffer from the disadvantages discussed above. Non-suspended diamond quantum photonic devices can be created using hybrid semiconductor-on-diamond structures \cite{Chakravarthi2022}, in which photons guided by a high refractive index optical waveguide layer interact evanescently with a supporting diamond substrate. Here we show that this hybrid platform can also be used to create non-suspended cavity optomechanical devices. Aided by diamond's high speed of sound, which exceeds that of all other materials, confinement of acoustic modes within the optical waveguide layer is possible. We present a gallium-phosphide (GaP) on diamond optomechanical cavity design that harnesses both the principle of total internal acoustic reflection \cite{Fu2019,Mayor2021} and suppression of mechanical loss through modal symmetries \cite{Zhang2021,Liu2022,Sarabalis2017} to realise devices that enhance optical coupling to low dissipation mechanical resonances, and show that these mechanical resonances can be coupled to spin qubits in the diamond substrate.

\begin{figure}[tb] 
\centering
\includegraphics[width=10cm]{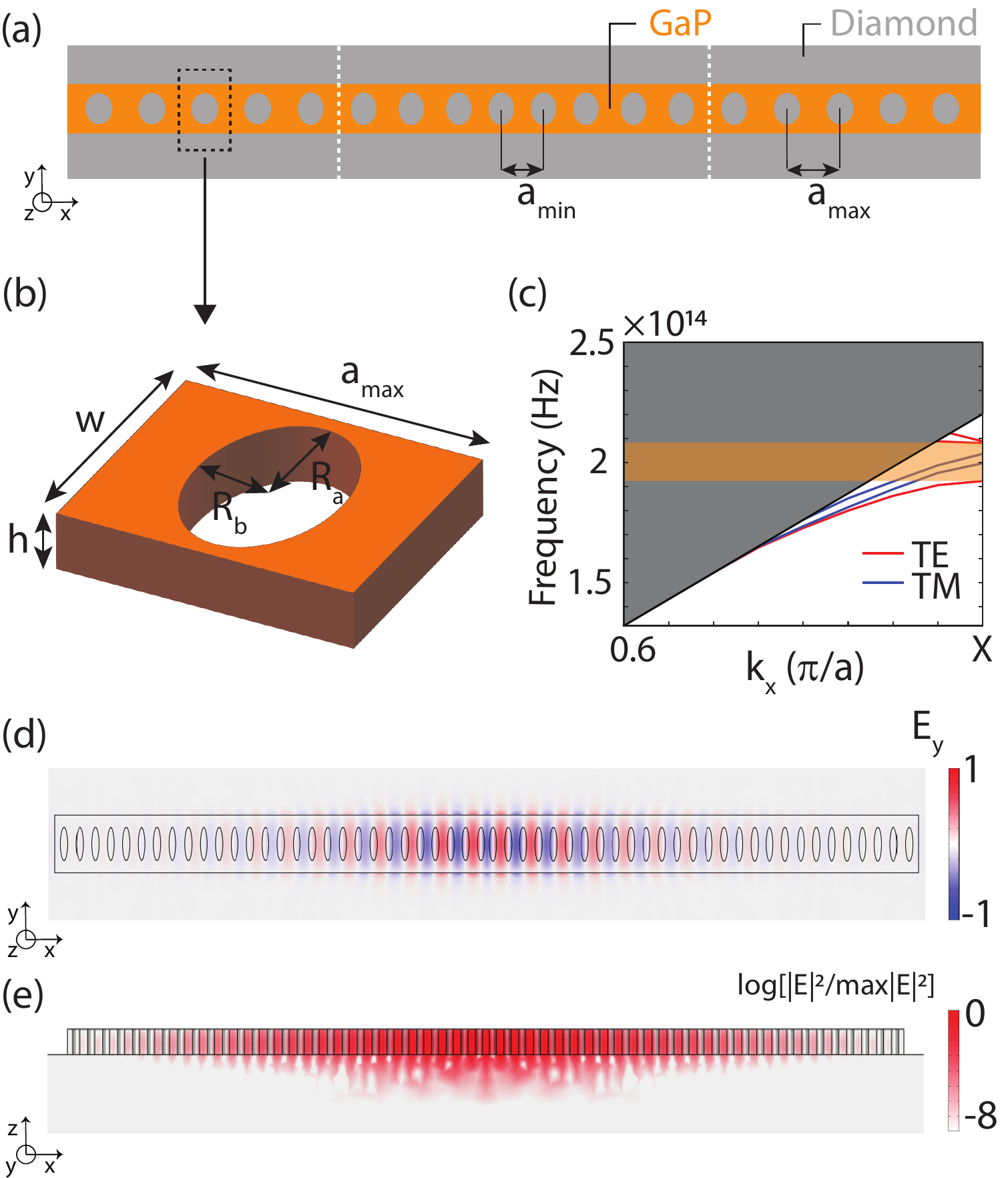}
\caption{(a) Schematic of the hybrid GaP-on-diamond optomechanical cavity. White dashed lines indicate the boundary between the cavity and mirror regions for this device. (b) Zoomed in schematic of a unit cell of the optomechanical crystal cavity. (c) Optical band structure of the device mirror region. The shaded area indicates an effective optical bandgap between two TE-like bands (red solid lines). (d) Optical cavity field profile of the fundamental TE-like cavity mode taken in the center plane of the GaP waveguide. (e) Vertical cross-section of the normalized electric field energy density of the TE-like cavity mode plotted on log-scale.}
\label{fig:fig1}
\end{figure}

Hybrid GaP-on-diamond devices have previously been used for optical waveguides \cite{Fu2008}, whispering gallery mode resonators \cite{Barclay2009, Gould2016}, and photonic crystal cavities \cite{Huang2021, Chakravarthi2022} that couple light to spin qubits in the diamond substrate. Optical confinement in these devices is possible since GaP's refractive index ($n_\text{GaP}=3.05$ at 1,550 nm) \cite{Khmelevskaia2021} is higher than diamond's ($n_\text{diamond}=2.39$ at 1,550 nm) \cite{Phillip1964}.
Similarly, since GaP's speed of sound is lower than that of diamond ($\sim 12,000$ m/s for transverse waves \cite{Flannery2003}) it is expected that GaP layer can support bound acoustic modes. However, the optomechanical properties of GaP-on-diamond structures have not been previously studied.

Here we show that by patterning the GaP layer with an optomechanical crystal cavity \cite{Eichenfield2009, Schneider2019}, both optical and acoustic modes can be localized within wavelength-scale volumes. In Sec.\ 2 we show that these modes can exhibit low loss, as characterized by optical and mechanical quality factors $Q_\text{o} > 10^4$ and $Q_\text{m}\sim10^3 - 10^6$, respectively, and optomechanical coupling that is sufficiently strong for coherent photon-phonon interactions. We find that a mechanical mode dependent trade-off between high $Q_\text{m}$ and high optomechanical coupling rate $g_\text{om}$ can be mitigated by harnessing symmetry dependent cancellation of mechanical radiation channels \cite{Zhang2021}, resulting in devices that simultaneously exhibit high $Q_\text{m}$ and $g_\text{om}$. Finally, in Sec.\ 3 we study the potential use of these devices for coherently coupling light to the mechanical resonator, and the mechanical resonator to electron spins in the diamond lattice.

\section{Device design}

The proposed device consists of an optomechanical crystal cavity formed by air holes in a GaP nanowire waveguide on a diamond substrate, as shown in Figs.\ 1(a) and 1(b). The cavity is created by quadratically tapering the hole spacing from $a_\text{max}$ to $a_\text{min}$. Bragg mirrors that reflect mechanical and optical fields at frequencies falling within their photonic and phononic bandgaps, respectively, are formed on either side of the cavity in the region of constant hole spacing. Within the region of tapered hole spacing, a cavity that supports co-localized optical and mechanical modes can be formed. Using finite element method simulations (COMSOL Multiphysics), devices were designed through tuning of six independent parameters--hole major and minor radii, $R_\text{a}$ and $R_\text{b}$, $a_\text{min}$ and $a_\text{max}$, and the nanobeam width, $w$, and height, $h$--to maximize figures of merit described below. 

\subsection{Optical cavity design}

The design procedure for the optical cavity follows Refs.\ \cite{Quan2011}. Optical bandstructures were first calculated for unit cells of varying hole size, cross-section, and lattice constant. A typical optical bandstructure for the mirror region unit cell is shown in Fig.\ 1(c). 
Waveguide modes lying below the diamond light-line are guided and form the basis for optical cavity modes when the tapered defect is introduced. 
They can be classified into TE-like (electric field dominantly polarized in lateral $y$-direction) or TM-like (electric field dominantly polarized in the vertical $z$-direction). An effective optical bandgap is formed in the shaded yellow region between TE-like optical bands. 
To create a cavity, the hole spacing is reduced, raising the energy of all of the bands. This shifts the $X$-point (in-plane wavevector $k_x = \pi/a$) TE-like valence bandedge of the cavity region unit cells into the bandgap of the surrounding mirror regions. The lattice constant tapering creates an effective potential-well for bandedge waveguide modes along the $x$-direction that provides the optical confinement of the cavity modes studied here.

Parameters for two cavities, labeled Devices A and B, are listed in Table 1.  Device A and B are optimized for photon-phonon and spin-phonon coupling, respectively, using different mechanical modes, as discussed in Sec.\ 2.2. 
Figure 1(d) shows the resulting optical field profile of the fundamental TE-like optical cavity mode formed using Device A's geometry with a cavity region tapered over 8 lattice constants and 25 sets of mirror holes. The field is concentrated in the GaP, as expected for a valence band mode \cite{joannopoulos2008photonic}. As summarized in Table 2, it has frequency $\omega_\text{o}/2\pi = 194\,\text{THz}$ and $Q_\text{o} = 23,000$. Optical loss is limited primarily by leakage into the diamond substrate, which could be reduced using a diamond ridge structure as in Ref.\ \cite{Barclay2009b} or employing a tapered waveguide width as in Ref.\ \cite{Zhang2021, Zhang2022}. Device B supports an optical mode with similar properties, also summarized in Table 2.

\begin{table}[!tb]
\centering
\caption{Dimensions used for devices A and B.}
\begin{tabular}{ccc}
\hline
Parameters  &  \makecell{Cavity optomechanics \\ optimized\\ \small{(Device A)} } & \makecell{Spin-optomechanics \\ optimized\\ \small{(Device B)} }\\
\hline
$w$ & 1143 nm & 1060 nm \\
$h$ & 431 nm & 450 nm \\
$R_\text{a}$ & 340 nm & 300 nm \\
$R_\text{b}$ & 70 nm & 62 nm\\
$a_\text{max}$ & 297 nm & 287 nm\\
$a_\text{min}$ & 291 nm & 280 nm\\
\hline
\end{tabular}
\label{tab:shape-functions}
\end{table}

\subsection{Mechanical properties and optomechanical coupling}
 
\begin{figure}[!tb]
\centering
\includegraphics[width=10cm]{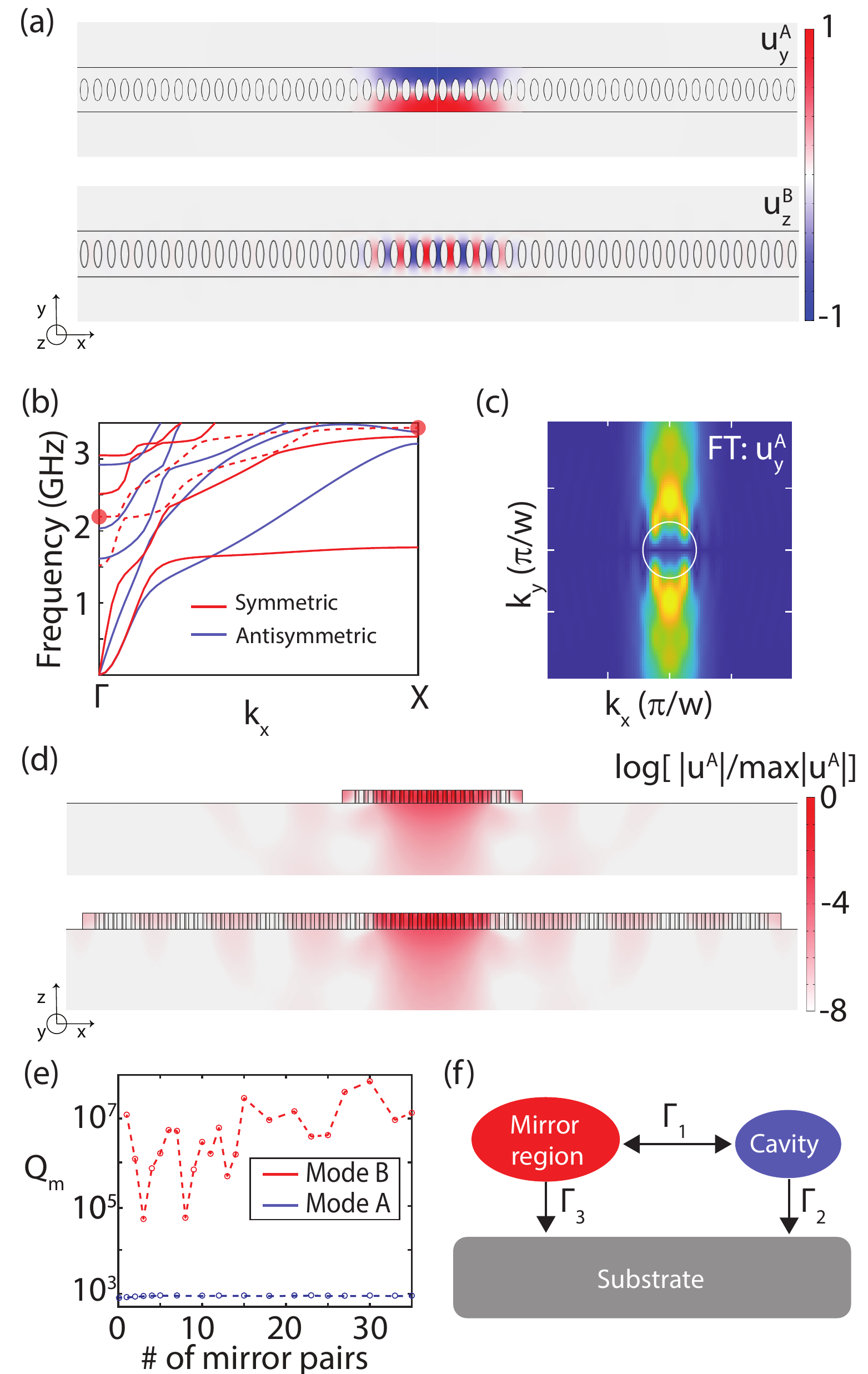}
\caption{(a) Normalized displacement profile ($y$-component) of mechanical mode $\mathcal{M}_\text{A}$ in the center plane of the GaP waveguide. (b) phononic band structure of the mirror region unit cell. The red solid points denote the resonance frequency of the mechanical defect mode. (c) Fourier transform of $u_\text{y}^{\text{A}}$ at the GaP-diamond interface. The white circle denotes the dispersion of a transverse acoustic mode in diamond. Here $d=6,000$ nm. (d) Normalized mechanical displacement field ($u$) for (top) zero mirror pairs and (bottom) 20 mirror pairs placed on both sides of the cavity region. (e) Simulated values of $Q_\text{m}$ for varying number of mirror hole pairs. Black and blue lines indicate mechanical modes $\mathcal{M}_\text{A}$ and $\mathcal{M}_\text{B}$, respectively. (f) Mechanical loss channels of a non-suspended mechanical crystal cavity. }
\label{fig:fig2}
\end{figure}

The cavity designs described above also support localized mechanical modes, two of which we analyze here: mode $\mathcal{M}_\text{A}$ is a mechanical breathing mode with large photon-phonon and spin-phonon coupling, and mode $\mathcal{M}_\text{B}$ is a Rayleigh mode with low mechanical dissipation and large spin-phonon coupling. Their displacement profiles are shown in Figs.\ 2(a). Mode $\mathcal{M}_\text{A}$ has frequency $\Omega_\text{m}^\text{A}/2\pi=$ 2.15 GHz and possesses high optomechanical coupling to the fundamental TE-like optical cavity mode presented above. If we qualitatively represent the GaP nanobeam as two rails connected by cross-ties, mode $\mathcal{M}_\text{A}$'s motion is described by opposing $y$-polarized displacement $u_\text{y}$ of the rails. Mode $\mathcal{M}_\text{B}$ has frequency $\Omega_\text{m}^\text{B}/2\pi=$ 3.30 GHz, and its dominant displacement is $z$-polarized. As with the photonic mode analyzed above, both mechanical modes are localized within the defect region due the variation of the phononic bandstructure within the cavity. 

Insight into the properties of both modes can be obtained from the phononic bandstructure of the mirror region unit cell, shown in Fig.\ 2(b). Mode $\mathcal{M}_\text{A}$ is formed from the $\Gamma$-point ($k_x = 0$) phononic crystal waveguide mode indicated in Fig.\ 2(b).  $\Gamma$-point modes are periodic, and as a result, each unit cell's contribution to optomechanical coupling, defined by the shift in optical resonance frequency per unit of mechanical displacement, add constructively. Note that this mode's dominant Fourier component is at $k_x = 0$, in contrast to the $\Gamma$-point mode with a large Fourier component at $k_x = 2\pi/a$ in Ref.\ \cite{Qi2021}. Mode $\mathcal{M}_\text{A}$'s per-phonon optomechanical coupling rate to the fundamental TE-like optical mode presented above is $g_\text{0}^\text{A}/2\pi=$ 110 kHz, where contributions from the photo-elastic effect (106 kHz) are much larger than contributions from moving boundaries (4 kHz). The is attributed to the relatively large nanobeam width, which was necessary to maintain high $Q_\text{o}$, and the valence band optical mode, which was chosen to maximize its separation from the diamond lightline but whose field is concentrated in the GaP regions between the holes. Both of these design characteristics reduce overlap of the optical field with the moving boundaries of the mechanical breathing mode. Although $\Gamma$-point modes are leaky due to being above the diamond sound-line, resulting in mechanical radiation into the diamond substrate, mode $\mathcal{M}_\text{A}$ has a predicted mechanical quality factor $Q_\text{m}^\text{A} = 900$. 
 Mode $\mathcal{M}_\text{A}$'s relatively low loss can be attributed to its odd symmetry in the $y$-direction, which results in a large fraction of its Fourier components falling outside of the diamond sound-cone, as shown in Fig.\ 2(c), despite its slow variation along the $x$-direction. This effect can also be described qualitatively by the cancellation of radiation from the out-of-phase motion in the $y$-direction of each side of the nanobeam, in analogy with the dipole cancellation presented in Ref.\ \cite{Zhang2022}.  Here the reported $Q_\text{m}^\text{A}$ and $g_\text{0}^\text{A}$ values were obtained after optimization of the device geometry (given in Table 1) to maximize the device's single photon optomechanical cooperativity, which is proportional to $ Q_\text{m}^\text{A} Q_\text{o}^\text{A} (g_\text{0}^\text{A})^2$ \cite{Safavi2019}.  As with the optical mode, it may be possible to further enhance $Q_\text{m}$ by creating a ridge on the diamond substrate to restrict leakage.
 
Mode $\mathcal{M}_\text{B}$ is formed from the $X$-point ($k_x = \pi/a$) photonic crystal waveguide mode indicated in Fig.\ 2(b). As a result, its displacement oscillates along the $x$-axis. This causes contributions to the optomechanical coupling from neighbouring unit cells to partially cancel, leading to a small $g_\text{0}^\text{B}/2\pi = 5.8$ kHz. However, as the $X$-point bandedge is below the diamond sound-line, mode $\mathcal{M}_\text{B}$ has low leakage into the substrate, resulting in high $Q_\text{m} \sim 10^6$. Note that Device B's geometry was optimized for strain coupling discussed below in Sec.\ 3.1, and that it may be possible to modestly increase $g_\text{0}^\text{B}$ through changes in device parameters.

Neither mode $\mathcal{M}_\text{A}$ or $\mathcal{M}_\text{B}$ fall within a full phononic bandgap of the mirror region, as shown by the phononic bandstructure in Fig.\ 2(b). This is in part due to the presence of the diamond substrate, which breaks the device's vertical symmetry, allowing otherwise forbidden coupling between modes with different parity in the $z$-direction  \cite{Chan2012, Eichenfield2009}. However, the gradual nature of the cavity's tapered defect minimizes coupling to modes away from the bandedges from which each cavity mode is formed, enabling the relatively high $Q_\text{m}$ of the cavity modes studied here.

\begin{table*}[!t]
\centering
\caption{\bf Simulated values of the key figures of merit for devices A and B.}
\resizebox{\textwidth}{!}{
\begin{tabular}{@{}c|cc|cccc|cccc@{}}

  Device & $\omega_\text{o}/2\pi$ & $Q_\text{o}$ & $\Omega_\text{m}/2\pi$ & $Q_\text{m}$ & $m_\text{eff}$ & \makecell{Mech. mode\\profile}& $g_\text{0}/2\pi$ & $\epsilon_\text{zpm} (\frac{\text{strain}}{\text{phonon}})$ & \makecell{per-photon\\$C_\text{om}$} & \makecell{per-phonon\\$C_{\text{sm,SiV}^-}$}\\
\hline
A  & 194.1 THz & 23,000  & 2.15 GHz & 900 & 563 fg & Breathing & 110 kHz &  $0.7\times10^{-9}$ & $2.4\times10^{-6}$ &  0.008\\
B  & 198.6 THz & 29,800  & 3.31 GHz & $>10^6$ & 301 fg & Rayleigh & 5.8 kHz & $1.3\times10^{-9}$ & $6.1\times10^{-6}$ &  20\\

\end{tabular}}
\end{table*}

Insight into the loss mechanisms limiting each mode's mechanical dissipation can be obtained from calculating $Q_\text{m}$ as a function of the number of mirror hole pairs, as shown in Fig.\ 2(e). We see that mode $\mathcal{M}_\text{A}$'s mechanical quality factor is not significantly affected by the mirror region, indicating that phononic radiation into the substrate dominates its mechanical loss. This is evident from the simulated displacement field shown in Fig.\ 2(d) for two different mirror lengths. In contrast, mode $\mathcal{M}_\text{B}$'s mechanical quality factor tends to increase with the number of mirror hole pairs. This indicates that loss into diamond substrate plays a smaller role than for mode $\mathcal{M}_\text{A}$. Oscillations in $Q_\text{m}^\text{B}$ for a small number of mirror hole pairs can be qualitatively explained by the model illustrated in Fig.\ 2(f), which is adapted from Ref.\ \cite{Eichenfield2009}. In this model, mechanical loss channel $\Gamma_1$ describes coupling between the localized mode and equal-frequency standing wave modes of the mirror region. The physical contact between the diamond substrate and the nanobeam introduces additional loss channels $\Gamma_\text{2}$ and $\Gamma_\text{3}$, corresponding to mechanical radiation loss from the cavity mode into the substrate and radiation loss from the mirror region into the substrate, respectively. As the number of the mirror pairs changes, mechanical mode $\mathcal{M}_\text{B}$ couples to a varying spectrum of standing wave modes of the mirror, whose free spectral range and isolation from the substrate decreases as the mirror region becomes larger.

\section{Coherent mechanical coupling to spins and photons}
The device presented here can enable coherent coupling between phonons and both photons and electron spins of diamond colour centers. This can enable optical control of electron spin qubits \cite{atature2018material} without relying upon their intrinsic optical transitions \cite{Shandilya2021, Wang2020}.  In this section we discuss the feasibility of realising these coherent interfaces with the device presented above. 

\begin{figure}[!t]
\centering
\includegraphics[width=10cm]{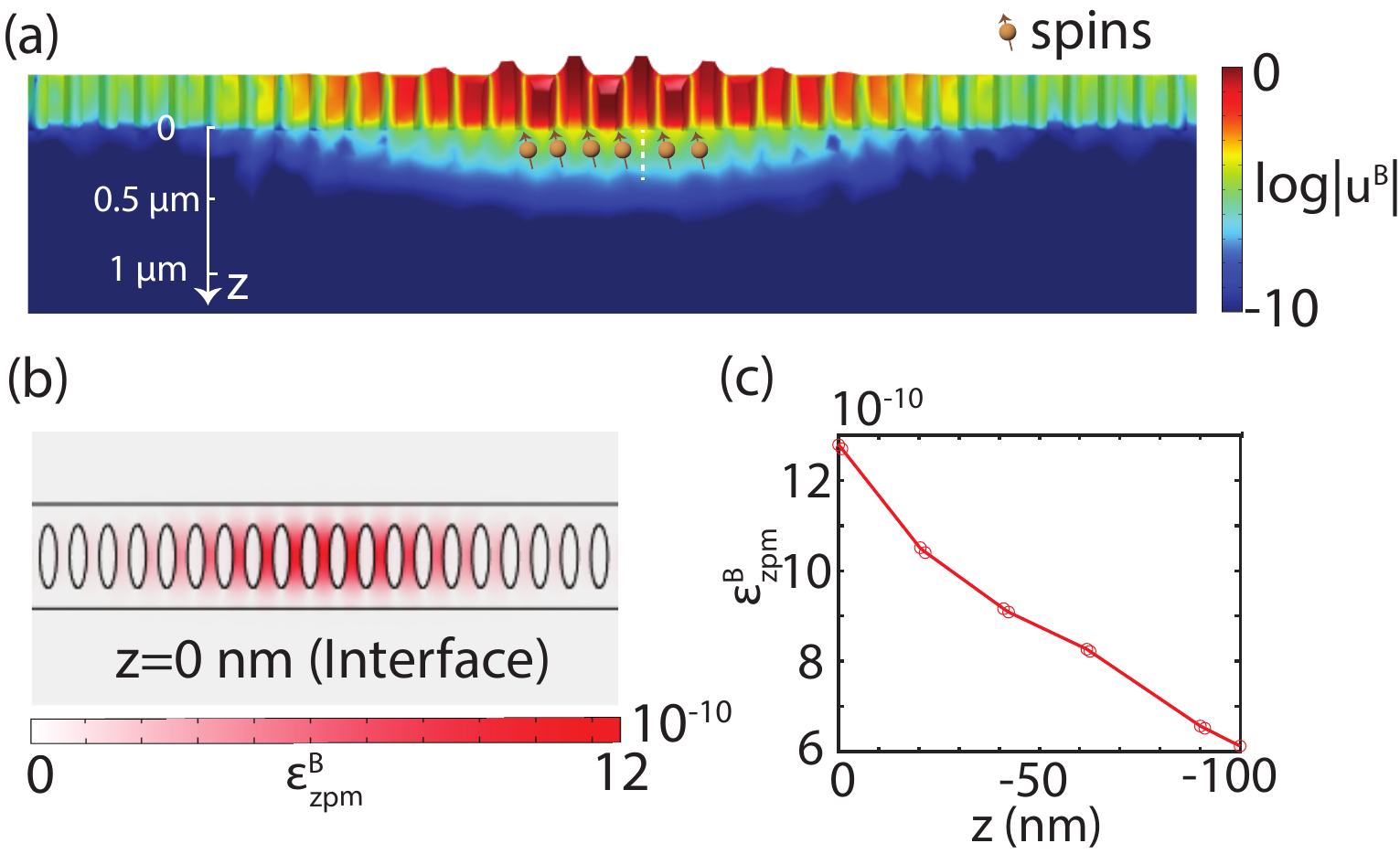}
\caption{(a) Cross-section of the displacement of mode  $\mathcal{M}_\text{B}$, with the position of diamond color centre spins that could be coupled the mechanical mode shown for illustrative purpose. A logarithmic scale is used in the figure. (b) Simulated per-phonon strain at the GaP-diamond interface. In both (a) and (b) the total strain is estimated by $\sqrt{S_{xx}^2+S_{yy}^2+S_{zz}^2}$, where $S_{ii}$ are the axial strain components. (c) The maximum strain field with respect to the depth below the diamond surface.}
\label{fig:fig3}
\end{figure}

\subsection{Spin-phonon coupling}

Coupling between mechanical resonators and electron spins of defects in diamond such as the silicon-vacancy center (SiV) and the nitrogen-vacancy center (NV) can be induced by the oscillating strain in the diamond crystal lattice created by resonator vibrations \cite{macquarrie2013mechanical, Lee2017}. The strain field modifies the wavefunctions of electron spin states localized to defects in the crystals lattice, and can coherently drive population between spin states when the mechanical mode frequency is resonant with a spin transition \cite{Lee2017}. 

Creating a coherent spin-phonon interface requires a spin-phonon cooperativity $C_\text{sm} = 4g_\text{sm}^2/(\gamma_\text{m}\gamma_\text{s}) > 1$, where $\gamma_\text{m} = \Omega_\text{m}/Q_\text{m}$ is the mechanical mode's energy dissipation rate, $\gamma_\text{s}$ is the spin dephasing rate, and $g_\text{sm}$ is the spin-phonon coupling rate \cite{Wang2020}. This coupling rate depends on the strain generated in the diamond crystal by the displacement field from by a single phonon, $\epsilon_\text{zpm}$, as well as the strain susceptibility of the spin system of interest. As shown in Table 2, both modes $\mathcal{M}_\text{A}$ and $\mathcal{M}_\text{B}$ create a maximum per-phonon strain in the diamond substrate of $\sim 10^{-9}$, with $\mathcal{M}_\text{B}$ having a strain field that is larger than that of $\mathcal{M}_\text{A}$, and approximately twice are large as that experimentally demonstrated in a diamond microdisk spin-optomechanical system \cite{Shandilya2021}. This value is smaller than what may be achievable in suspended diamond optomechanical crystal cavities \cite{Cady2019}, in part due to the diamond being located below the cavity structure rather than directly within it. Mode $\mathcal{M}_\text{B}$'s predominantly $z$-polarized motion (see Fig.\ 3(a)) shares similarities with layered Rayleigh waves used for driving spins of diamond SiV center \cite{Maity2020} and NV center \cite{Golter2016}. Mode $\mathcal{M}_\text{B}$'s strain-per-phonon at the GaP-diamond interface is plotted in Fig.\ 3(b), and its variation with depth in the diamond substrate is shown in Fig.\ 3(c); it falls to $\epsilon_\text{zpm}|_{z = 100\text{nm}} = 0.6 \times 10^{-9}$ at $z = 100\,\text{nm}$ below the surface.

A key feature of Mode $\mathcal{M}_\text{B}$ that makes it suited for spin-phonon coupling is its high $Q_\text{m}^\text{B} > 10^6$. If we assume $Q_\text{m}^\text{B} = 10^6$, we estimate $C_\text{sm,SiV$^-$} = 20$  for an SiV electron spin located 100 nm below the diamond surface. This assumes $\gamma_\text{s}/2\pi=1$ MHz, as observed for SiV spins at mK temperatures \cite{Meesala2018}, and utilizes the relationship $g_{\text{sm,SiV}^-}/2\pi=d_{\text{SiV}^-}\epsilon_\text{zpm}=130$ kHz/phonon where $d_{\text{SiV}^-}=0.1$ PHz/strain \cite{Meesala2018} is the spin-strain susceptibility for the electronic ground states of negatively charged SiV colour centers. In contrast, $\mathcal{M}_\text{A}$'s lower $Q_\text{m}$ is expected to prevent it from reaching the coherent spin-phonon coupling regime.

\subsection{Optomechanical coupling}

Coherently coupling the optomechanical cavity to light is necessary  for connecting the spin-phonon interface described above to photons \cite{Shandilya2021}, and more generally, for applications in classical and quantum information processing \cite{Safavi2019}. This requires optomechanical cooperativity $C_\text{om}=4N_\text{cav}g_\text{0}^2/(\gamma_\text{m} \kappa) > 1$, where $\kappa = \omega_\text{o}/Q_\text{o}$. Satisfying this criteria is greatly aided by its dependence on $N_\text{cav}$, the intracavity photon number, which parametrically enhances the optomechanical coupling between a single phonon and a single photon \cite{Aspelmeyer2014}. Multi-photon $C_\text{om}>1$ has been routinely demonstrated in 1D suspended OMC cavity systems in a variety of materials including GaP \cite{Schneider2019} and diamond \cite{Burek2016}, as well as in diamond microdisks \cite{Lake2018, Shandilya2021}. Despite the relatively low $g_\text{om}^\text{B}$ of mode  $\mathcal{M}_\text{B}$, it is capable of operating with $C_\text{om}>1$ for $N_\text{cav} > 1.7\times10^5$ photons input to the cavity, assuming $Q_\text{m} = 10^6$. Mode $\mathcal{M}_\text{A}$, which has a higher $g_\text{om}$ but lower maximum $Q_\text{m}$, requires approximately $N_\text{cav} > 4.3\times10^5$ photons to reach $C_\text{om} > 1$.
Similar or higher photon numbers have been demonstrated in suspended diamond optomechanical crystals \cite{Burek2016} and microdisks \cite{Lake2018, Shandilya2021}, limited primarily by heating and thermo-optic dispersion of the cavity \cite{Lake2018}. It is is expected that the improved heat-sinking of the GaP-on-diamond structure will allow access to higher $N_\text{cav}$ than previously demonstrated.  

Improvements to $Q_\text{m}$ and $Q_\text{o}$, for example through patterning a ridge in the diamond substrate \cite{Barclay2009b}, would further aid in operating in the regime of coherent spin- and photon-phonon coupling.
Increasing $Q_\text{o}$ would also be beneficial for allowing the device to operate deeper in the sideband resolved regime from its current operating point of $\Omega_\text{m}^\text{B}/\kappa_\text{B} \sim 0.5$.
Improvement to device performance may also be possible by designing it to operate at visible wavelengths: the higher refractive index of GaP ($n\sim 3.4$ at 637 nm) would increase $Q_\text{o}$, while the smaller device dimensions and resulting tighter photon confinement will increase $g_\text{om}$. Investigating these effects, and the corresponding changes to the mechanical properties of the device, requires additional studies beyond the scope of this work.

\section{Conclusion}
In summary, we have designed a non-suspended optomechanical crystal cavity patterned in a GaP nanobeam on a diamond substrate that will enable coherent photon-phonon and spin-phonon interactions. The interaction between a high quality factor TE-like optical cavity mode and two mechanical cavity modes with distinct properties were studied. A mechanical breathing mode was found to exhibit strong optomechanical coupling to the optical mode with optomechanical coupling rate $g_\text{0}/2\pi=110.0$ kHz while maintaining a mechanical quality factor of $\sim 900$. A Rayleigh-like cavity mode with a smaller optomechanical coupling is shown to be of promise for coherent spin-phonon coupling owing to its high $Q_\text{m}>10^6$, which results in part from the high speed of sound of the diamond substrate. 

For the predicted device performance, this optomechanical cavity can be used to create a coherent optomechanical interface between electron spins in the diamond lattice and photons at telecommunication wavelength. This spin-optomechanical platform could be integrated with phononic waveguides and superconducting circuits in order to realize recently proposed quantum transducers \cite{neuman2021phononic}. Such technologies may benefit from the large electro-optic and piezo-electronic coefficients of GaP, which opens to the door to directly interfacing both its optical and mechanical modes with microwaves \cite{sahu2023entangling}.

\end{document}